\documentclass[aps,prl,reprint,superscriptaddress,nofootinbib]{revtex4-2}
\usepackage{amsmath,amssymb,bm,mathtools}
\usepackage{graphicx}
\usepackage{xcolor}
\usepackage{microtype}
\usepackage{comment}
\usepackage{hyperref}
\hypersetup{colorlinks=true,linkcolor=blue!55!black,citecolor=blue!55!black,urlcolor=blue!55!black}

\newcommand{\dif}{\mathrm{d}}

\begin{document}

\title{Phase Memory Controls Radiative Depolarization Near an Isolated Spin Resonance}

% Author list and affiliations are placeholders for the first manuscript draft.
\author{Zhe Duan}
\thanks{Contact author: duanz@ihep.ac.cn}
\affiliation{Institute of High Energy Physics, Chinese Academy of Sciences, Beijing 100049, China}
% \author{...}
% \affiliation{...}

\date{October 13, 2026}

\begin{abstract}

Radiative depolarization near spin resonances in electron
storage rings is conventionally described by correlated-sideband
and local resonance-density theories, but a quantitative
description of the crossover between them has remained lacking.
Here we derive a unified finite-memory response from
stochastic spin dynamics near an isolated resonance.
By separating the coherent resonance contribution from
stochastic fluctuations, we show that the connected correlation
of the accumulated resonance-relative spin phase determines
the leading off-center weak-resonance depolarization rate.
Both conventional theories emerge as controlled asymptotic
limits of this response, which also captures the surviving synchrotron phase memory governing the crossover.
Monte Carlo spin tracking validates the predicted rates
throughout the crossover and reproduces the recurrence-induced
generalized-sideband structure.
More broadly, the connected-correlation formulation
provides a basis for extending radiative depolarization theory to more general stochastic spin-orbit dynamics in storage rings.

\end{abstract}

\maketitle

Beam polarization is a powerful tool for precision physics at
high-energy circular $e^+e^-$ colliders, notably enabling precision
beam-energy calibration through resonant depolarization
~\cite{Arnaudon1995,BlondelGianfelice2021}.
In electron storage rings, synchrotron radiation both drives
Sokolov--Ternov self-polarization through spin-flip emission
~\cite{SokolovTernov1964} and produces radiative spin diffusion through
stochastic orbital recoil~\cite{BaierOrlov1966}.
Their competition determines the equilibrium polarization and its
buildup time~\cite{DerbenevKondratenko1973,Mane1987}.
The resulting radiative depolarization can be strongly enhanced near spin
resonances, where spin motion becomes highly sensitive to
radiation-induced orbital fluctuations.

Near a spin resonance, radiative depolarization reflects the interplay
of the spin-resonance driving term, synchrotron motion, and stochastic
photon emission.  Synchrotron motion periodically modulates the instantaneous resonance
detuning and hence the accumulated resonance-relative spin phase,
while photon recoil perturbs the phase coherence between
resonance-driven spin responses separated in time.
Two limiting dynamical regimes follow from whether this coherence
survives.  When radiation-induced decorrelation remains weak over
successive synchrotron periods, the resonance-driven spin responses
remain phase coherent. 
In the correlated regime, the conventional
weak-radiation description expresses depolarization
through the energy gradient of the invariant spin
field (ISF)~\cite{BarberEllisonHeinemann2004},
with synchrotron modulation producing higher-order
sidebands~\cite{DKS1979,Yokoya1983Chromaticity}.

In the opposite limit, radiation destroys the phase coherence between
successive resonance passages, so that their contributions to spin
diffusion add incoherently.  For sufficiently rapid passages, the
depolarizing contribution of each passage, after averaging over the
randomized spin phase, scales inversely with the crossing speed
~\cite{FroissartStora1960,Kondratenko1974}, whereas the passage flux
scales proportionally to it.  The crossing speed therefore cancels,
leaving a depolarization rate governed by the instantaneous density on
the resonance surface~\cite{Kondratenko1974,DK1975}.
This uncorrelated regime becomes the relevant asymptotic limit at
sufficiently high energies~\cite{DKS1979}.

The two limits therefore represent distinct reductions of the same
radiative spin dynamics, distinguished by the survival of phase
coherence between spin responses separated in time.  Between these limits lies an intermediate regime
in which radiation progressively erodes, but has
not yet eliminated, the phase memory.

Synchrotron sidebands were directly observed at LEP, while the
depolarizing behavior measured up to 60.6~GeV was consistent with the
correlated higher-order sideband theory
~\cite{Assmann1999,Assmann2001}.
At energies where uncorrelated resonance passages were expected to
become relevant, however, the polarization was too small for a
quantitative test~\cite{Assmann2001}.

More recently, Monte Carlo spin tracking for CEPC~\cite{Xia2023}
revealed a broad crossover between the correlated and uncorrelated
limits, with an intermediate region described quantitatively by
neither limiting theory.
A quantitative dynamical response connecting these
limiting descriptions across the finite-memory
crossover has remained lacking.

Here we derive a finite-memory response from
the connected correlation of the accumulated
resonance-relative spin phase near an isolated
resonance subject to stochastic synchrotron motion.
The correlated ISF-gradient and uncorrelated
resonance-density descriptions emerge as
controlled asymptotic limits, while direct
stochastic spin tracking validates the
finite-memory prediction throughout the crossover.

For an electron circulating in a storage ring, the spin vector
$\bm S$ evolves according to the
Thomas--Bargmann--Michel--Telegdi (T--BMT) equation
~\cite{Thomas1926,BargmannMichelTelegdi1959},
$d\bm S/d\theta=\bm\Omega(\bm z,\theta)\times\bm S$, where
$\theta$ is the ring azimuth, $\bm z$ denotes the orbital phase-space
coordinates, and $\bm\Omega$ is the spin-precession vector.
On the closed orbit, the spin motion is characterized by the periodic unit
spin solution $\bm n_0(\theta)$ and the closed-orbit spin tune $\nu_0$.
Away from the closed orbit, these generalize to the ISF
$\bm n(\bm z,\theta)$ and an amplitude-dependent spin tune
~\cite{BarberEllisonHeinemann2004}.
Spin resonances arise when a harmonic of the spin-driving perturbation
becomes phase synchronous with the spin precession.
Examples include integer resonances ($\nu_0=k$) and first-order
intrinsic resonances ($\nu_0\pm\nu_y=k$), where $k\in\mathbb Z$
and $\nu_y$ is the vertical betatron tune.

To isolate the effects of synchrotron motion and radiation on
resonance-relative phase coherence,
we adopt the standard
single-resonance model~\cite{Mane2005}, 
retaining a fixed effective
resonance strength while allowing stochastic longitudinal motion
to modulate the resonance detuning.
Similar single-resonance spin-tracking models incorporating
stochastic synchrotron motion have been used to study
resonant depolarization at FCC-ee~\cite{Koop2018}.
Let $\epsilon$ denote the effective strength of the
retained resonance harmonic and $\nu_{\rm res}$ its resonance tune.
With synchrotron tune $\nu_s$, relative energy deviation
$\delta_\epsilon$, and rms relative energy spread $\sigma_\epsilon$, we
define
$x=\delta_\epsilon/\sigma_\epsilon$ and
$\psi=\nu_s\theta$, so that one nominal synchrotron period is
$\Delta\psi=2\pi$.

The instantaneous resonance detuning $\Delta\nu_{\rm res}(\psi)$
measures the spin-precession rate relative to the retained harmonic,
expressed in tune units. 
To linear order in the energy deviation, its normalized form is
$\Delta(\psi)=\Delta\nu_{\rm res}(\psi)/\nu_s
=\delta+Mx(\psi)$,
where
$\delta=\frac{\nu_0-\nu_{\rm res}}{\nu_s}$,
$M=
 \frac{\sigma_\epsilon}{\nu_s}
 \left.
 \frac{\partial\Delta\nu_{\rm res}}
      {\partial\delta_\epsilon}
 \right|_0$ is the synchrotron modulation index.
For an integer spin resonance in a planar ring,
where $\nu_0\simeq a_e\gamma$ with $a_e$ the electron magnetic anomaly
and $\gamma$ the Lorentz factor,
$M\simeq\nu_0\sigma_\epsilon/\nu_s$.
The normalized resonance strength is $U={|\epsilon|}/{\nu_s}$.

In a frame
co-precessing with the retained resonance harmonic,
we choose the spin-coordinate axes $(\bm e_x,\bm e_y,\bm e_z)$ with
$\bm e_z$ along the unperturbed spin-precession axis and $\bm e_x$
along the spin-resonance driving term.
Hereafter, a dot denotes differentiation
with respect to $\psi$.
The single-resonance equation then becomes
\begin{equation}
\dot{\bm S}
=
\left[
U\bm e_x+\Delta(\psi)\bm e_z
\right]\times\bm S.
\label{eq:spin}
\end{equation}

To capture both periodic synchrotron recurrence and
radiation-induced phase decorrelation, we model the longitudinal
motion as a linear stochastic oscillator with radiation damping
and quantum excitation:
\begin{align}
 \dif x&=(-2ax-p)\dif\psi+\sqrt{4a}\,\dif W_\psi,
 \nonumber\\
 \dif p&=x\,\dif\psi,
 \label{eq:orbit}
\end{align}
where $p$ is the normalized conjugate synchrotron
coordinate, $a=\lambda_z/(2\pi\nu_s)$ with
$\lambda_z$ the longitudinal amplitude
damping decrement per turn, and $W_\psi$ is a
standard Wiener process. We consider $0<a<1$.
The stationary distribution is Gaussian with
$\langle x^2\rangle=\langle p^2\rangle=1$ and
$\langle xp\rangle=0$; angular brackets denote averages
over this stationary ensemble.
This model retains the essential phase-memory dynamics while
allowing the two-time correlations to be evaluated analytically.

Through the stochastic detuning $\Delta(\psi)$, the coupled
dynamics of Eqs.~(\ref{eq:spin}) and~(\ref{eq:orbit})
admits complementary frequency- and time-domain descriptions.
In the frequency domain,
synchrotron motion generates sidebands around the parent resonance;
in the time domain,
the response is governed by
the retention of resonance-relative phase coherence over finite lags.
We define the accumulated resonance-relative spin phase $\Phi(\psi)$
by
$\frac{\dif\Phi}{\dif\psi}=\Delta(\psi)$.
Using $\dif p/\dif\psi=x$, its increment over a dimensionless lag
$\vartheta$ is
\begin{equation}
 \Delta\Phi(\psi,\vartheta)
 \equiv
 \Phi(\psi+\vartheta)-\Phi(\psi)
 =
 \delta\vartheta+
 M[p(\psi+\vartheta)-p(\psi)] .
 \label{eq:phase}
\end{equation}
The corresponding single-trajectory unit phasor is
$q(\psi,\vartheta)=e^{i\Delta\Phi(\psi,\vartheta)}$.
Its stationary-ensemble average defines the full phase-correlation
function,
$G_{\rm full}(\vartheta)\equiv
\langle q(\psi,\vartheta)\rangle$,
which by stationarity depends only on $\vartheta$.

For the stationary Gaussian process, define the synchrotron-coordinate
autocorrelation
$C_p(\vartheta)=
\langle p(\psi+\vartheta)p(\psi)\rangle$.
Equation~(\ref{eq:phase}) and stationarity give
$\langle\Delta\Phi\rangle=\delta\vartheta$
and $\operatorname{Var}(\Delta\Phi)
 =2M^2[1-C_p(\vartheta)]$.
Since $\Delta\Phi$ is Gaussian, 
\begin{equation}
 G_{\rm full}(\vartheta) =
 e^{i\delta\vartheta}
 e^{-M^2[1-C_p(\vartheta)]}.
 \label{eq:gfull}
\end{equation}

The stochastic part of $\Delta\Phi$ is the endpoint difference of the
stationary coordinate $p$ and therefore has a bounded long-lag variance.
Accordingly, as $C_p(\vartheta)\to0$,
$e^{-i\delta\vartheta}G_{\rm full}(\vartheta)\to e^{-M^2}$.
This residual component arises from the coherent mean of the
synchrotron-modulated spin-resonance driving term.
A rotation of the spin frame about $\bm e_z$ by $Mp(\psi)$ removes $Mx$ from
the detuning term and transforms the resonance driving term to
$U[\cos(Mp)\bm e_x-\sin(Mp)\bm e_y]$.
Since
$\langle e^{-iMp}\rangle=e^{-M^2/2}$,
the transformed resonance driving contains a coherent mean component
of strength
$U_{\rm coh}=Ue^{-M^2/2}$,
whose mean-product correlation is $e^{-M^2}$.
Together with the mean detuning $\delta$, this coherent component
defines the reference spin motion, while the remaining zero-mean
fluctuations provide the stochastic driving responsible for radiative spin relaxation.

\begin{figure*}[t]
\centering
\includegraphics[width=0.95\textwidth]{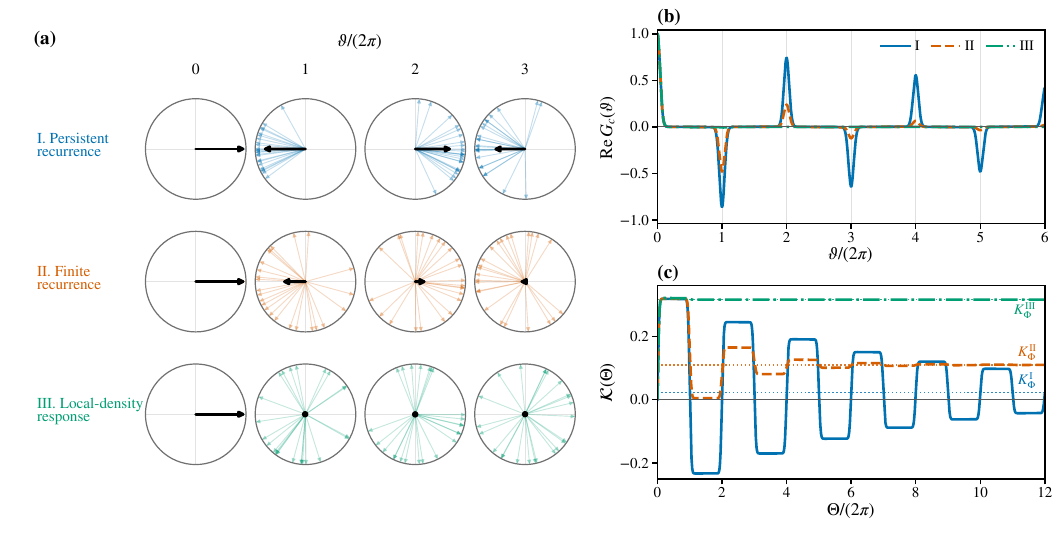}
\caption{
Resonance-relative phase memory and its contribution to radiative
depolarization.
(a) Ensembles of the single-trajectory unit phasor
$q=\exp(i\Delta\Phi)$ at $\vartheta/(2\pi)=0,1,2,3$ for cases I--III.
Faint vectors show representative stochastic realizations and the bold
vector denotes the full ensemble correlation
$G_{\rm full}=\langle q\rangle$.
(b) Corresponding connected phase-memory correlation
$\operatorname{Re}G_c(\vartheta)$, after subtraction of the coherent
mean product.
(c) Cumulative radiative response
$\mathcal K(\Theta)=
\operatorname{Re}\int_0^\Theta G_c(\vartheta)\,\dif\vartheta$;
horizontal lines denote
$K_\Phi=\mathcal K(\infty)$, evaluated from the full response.
Here $M=3.94$, $\delta=0.5$, and
$\Gamma=0.30$, $1.50$, and $11.94$ for I--III.
}
\label{fig:phase_memory}
\end{figure*}

We therefore remove
the mean-product contribution of the coherent component and define the connected phase-memory
correlation
\begin{equation}
 \begin{aligned}
 G_c(\vartheta)
 &\equiv
 G_{\rm full}(\vartheta)
 -e^{i\delta\vartheta}e^{-M^2}
 \\
 &=
 e^{i\delta\vartheta}
 \left[
 e^{-M^2[1-C_p(\vartheta)]}
 -e^{-M^2}
 \right],
 \\[1mm]
 K_\Phi(\delta,M,a)
 &\equiv
 \operatorname{Re}
 \int_0^\infty G_c(\vartheta)\,\dif\vartheta .
 \end{aligned}
 \label{eq:connected_kernel}
\end{equation}
Here ``connected'' has its standard statistical meaning: only the
mean-product contribution is subtracted, leaving the correlation of
the zero-mean fluctuations rather than introducing a phenomenological
decoherence prescription.
The connected response also admits a spectral interpretation:
$K_\Phi(\delta,M,a)=S_{\rm PM}(\delta;M,a)/2$,
where $S_{\rm PM}$ is the Fourier transform of
$e^{-i\delta\vartheta}G_c(\vartheta)$.
Thus the mean detuning selects the spectral component
of the fluctuating resonance-driving amplitude.

For Eq.~(\ref{eq:orbit}), the autocorrelation is
$C_p(\vartheta)
=e^{-a|\vartheta|}
[\cos(\omega|\vartheta|)
 +(a/\omega)\sin(\omega|\vartheta|)]$,
where $\omega=\sqrt{1-a^2}$.
Since $a>0$, $C_p$ decays exponentially, so
$G_c(\vartheta)\to0$ and the integral defining
$K_\Phi$ converges.
At the first nominal synchrotron recurrence,
$\vartheta=2\pi$, the weak-damping expansion gives
$\operatorname{Var}[\Delta\Phi(\psi,2\pi)]
\simeq4\pi aM^2$.
We denote this recurrence-loss parameter by
$\Gamma\equiv4\pi aM^2$, equivalent to the
conventional radiation-induced spin-phase diffusion
parameter~\cite{DKS1979,Yokoya1983SpinFlip,Blinov2022}.
It characterizes the radiation-induced loss of phase memory between successive synchrotron periods.
Because $\operatorname{Re}G_c$ is signed,
different recurrence lobes can contribute with
either sign to $K_\Phi$.

Figure~\ref{fig:phase_memory} visualizes how phase memory builds the
integrated response.  In (a), phasor ensembles sampled at successive
synchrotron periods refocus strongly in I, only partially in II, and
negligibly in III, illustrating the progressive loss of
resonance-relative phase coherence.  Panel (b) shows the corresponding
continuous connected correlation $\operatorname{Re}G_c(\vartheta)$,
and panel (c) its cumulative integral approaching $K_\Phi$.
The short-lag central lobe changes little from I to III, whereas the
later synchrotron recurrences progressively disappear.

The connection to depolarization follows from
the second-order response of the spin dynamics.
The fluctuating resonance-driving field first
generates a transverse spin response of order
$U$, whose feedback on the coherent-axis
polarization is of order $U^2$.
After averaging over the stationary orbital
ensemble, the projected response is governed
by $\operatorname{Re}G_c$.
The derivation is given in the End Matter.

Let $P(\psi)$ denote the slowly varying ensemble
polarization along the coherent reference spin axis.
After the initial memory transient, if $P$ varies negligibly over the
lags for which $G_c(\vartheta)$ contributes appreciably, its evolution
reduces to $dP/d\psi=-\rho_dP$, with
\begin{equation}
 \rho_d
 =
 U^2K_\Phi(\delta,M,a)+o(U^2).
 \label{eq:main}
\end{equation}
Here $\rho_d$ is the depolarization rate per unit synchrotron phase,
so that $2\pi\nu_s\rho_d$ is the depolarization decrement per turn.
Equation~(\ref{eq:main}) is the off-center weak-resonance result;
within narrow coherent resonance cores, coherent precession and
fluctuation-induced relaxation cannot in general be reduced to a
single scalar decay mode.

\begin{figure*}[t]
\centering
\includegraphics[width=0.95\textwidth]{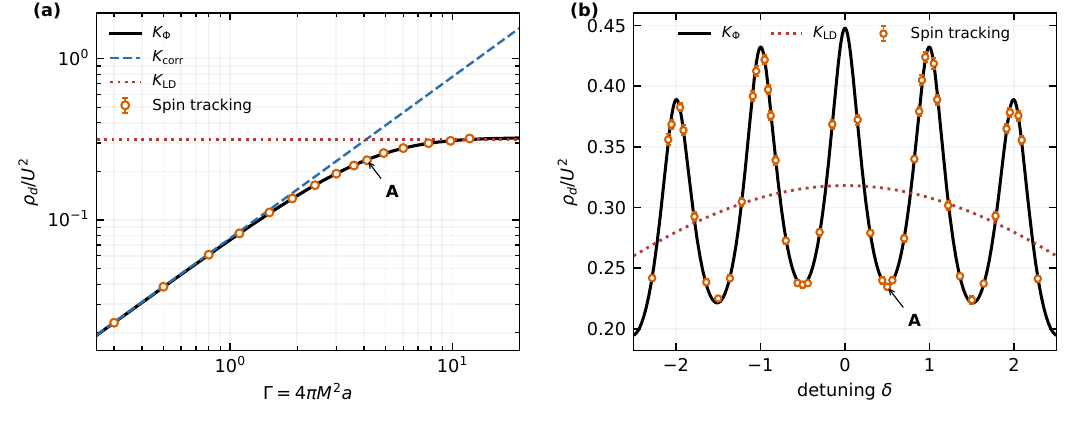}                                                                                                                                                                                                                                 
\caption{
Monte Carlo spin-tracking validation of the finite-memory
response.
(a) Normalized depolarization rate versus
$\Gamma=4\pi M^2a$ at $M=3.940$ and $\delta=0.5$.
Curves show $K_\Phi$ (solid), $K_{\rm corr}$ (dashed),
and $K_{\rm LD}$ (dotted); symbols denote spin tracking.
(b) Detuning scan at $M=3.940$ and $\Gamma=4.125$,
resolving the recurrence-induced generalized-sideband
structure predicted by $K_\Phi$ but absent from $K_{\rm LD}$.
All tracking uses $U=0.01$.
Point A in both panels denotes the common finite-memory
benchmark at $(\Gamma,\delta)=(4.125,0.5)$.
Error bars are statistical standard errors.
}
\label{fig:phase_memory_validation}
\end{figure*}

The relative contributions of the short-lag
response and subsequent synchrotron recurrences
in Fig.~\ref{fig:phase_memory} determine two
distinct asymptotic limits of $K_\Phi$.
When connected phase memory persists over successive synchrotron
periods, the weak-radiation expansion away from narrow coherent
sideband cores reduces to the correlated-sideband result.
In the local-density limit, the instantaneous detuning is effectively
constant over the short-lag response, while the integrated
contributions from subsequent recurrences become negligible, yielding
the local resonance-density result.
The two limits are
\begin{widetext}
\begin{equation}
\begin{aligned}
K_\Phi &\longrightarrow
K_{\rm corr}
=
2aM^2
\sum_{m=-\infty}^{\infty}
\frac{e^{-M^2}I_{|m|}(M^2)}
     {\bigl[(\delta+m)^2-1\bigr]^2},
&& \text{correlated-sideband limit},
\\[2mm]
K_\Phi &\longrightarrow
K_{\rm LD}
=
\pi p_\Delta(0)
=
\sqrt{\frac{\pi}{2}}\frac{1}{|M|}
\exp\!\left[-\frac{\delta^2}{2M^2}\right],
&& \text{local resonance-density limit}.
\end{aligned}
\label{eq:endpoints}
\end{equation}
\end{widetext}
Here $I_m$ is the modified Bessel function of the first kind,
and $p_\Delta$ is the stationary Gaussian density of the
instantaneous resonance detuning $\Delta=\delta+Mx$,
with mean $\delta$ and standard deviation $|M|$.
The resonance surface corresponds to $\Delta=0$.
The two limits reproduce, respectively, the weak-radiation
ISF-gradient result~\cite{Yokoya1983Chromaticity}
and the single-harmonic uncorrelated-passage result
~\cite{DKS1979,Kondratenko1974,DK1975}.
Their asymptotic derivations and required orderings
are given in the End Matter.

Thus $K_{\rm corr}$ and $K_{\rm LD}$ are limiting forms of the same
response $K_\Phi$, not additive depolarization channels.
Crucially, the local-density reduction requires both local freezing of
the instantaneous detuning over the short-lag response and suppression
of the integrated contributions from subsequent synchrotron
recurrences.  Large $|M|$ narrows the short-lag response to a scale
$\vartheta\sim |M|^{-1}$ but does not by itself suppress the recurrences.
When either condition is only partially satisfied, the full
$K_\Phi(\delta,M,a)$ is required.

We validate Eq.~(\ref{eq:main}) by Monte Carlo tracking
of Eqs.~(\ref{eq:spin}) and~(\ref{eq:orbit}),
using exact finite-step orbital propagation and
norm-preserving $SO(3)$ spin transport.
Depolarization rates are extracted from the late-time
exponential decay of the deterministic-ISF polarization
projection, with numerical details and convergence tests
given in the End Matter.

At fixed $M$, increasing $\Gamma$ progressively suppresses
synchrotron recurrences, and the tracking results in
Fig.~\ref{fig:phase_memory_validation}(a) follow the
finite-memory prediction $K_\Phi$ throughout the crossover.
Point A provides a stringent finite-memory benchmark:
the short-lag response is already localized, while
substantial recurrence survives.
Here $K_{\rm corr}$ and $K_{\rm LD}$ overestimate
$K_\Phi=0.236$ by $34.8\%$ and $33.6\%$, respectively,
whereas tracking agrees with the finite-memory prediction.

Figure~\ref{fig:phase_memory_validation}(b) provides an
independent detuning-domain test, reproducing the
recurrence-induced generalized-sideband peaks and valleys
absent from the smooth $K_{\rm LD}$ prediction.
These spectral features reflect surviving synchrotron
phase memory.
Across both panels, the maximum relative
deviation of the 59 tracking rates from
$K_\Phi$ is below $1.8\%$.

\begin{figure}[!htb]
\centering
\includegraphics[width=0.95\columnwidth]{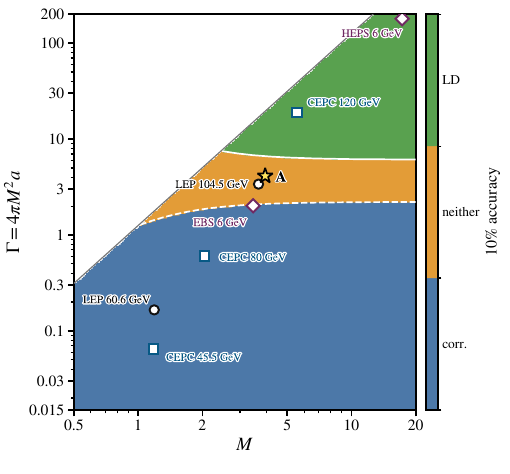}
\caption{
Validity map of the limiting depolarization theories
at $\delta=0.5$.
Blue and green indicate where $K_{\rm corr}$ and
$K_{\rm LD}$, respectively, agree with $K_\Phi$
within $10\%$; orange indicates where neither does.
White curves denote the $10\%$ accuracy boundaries,
not phase boundaries.
Markers represent selected storage-ring parameters
listed in Table~\ref{tab:machine_mapping} in the End Matter;
the star denotes tracking benchmark A of
Fig.~\ref{fig:phase_memory_validation}.
The uncolored region ($a\geq0.1$) is excluded
from the weak-damping map.
}
\label{fig:regime_map}
\end{figure}

Within the single-resonance model, the normalized
weak-coupling response depends only on
$(\delta,M,\Gamma)$, allowing different storage
rings to be compared through these dimensionless
parameters.
Figure~\ref{fig:regime_map} maps the crossover on the $(M,\Gamma)$
plane, with $a=\Gamma/(4\pi M^2)$,
at the representative inter-sideband detuning $\delta=0.5$.
The $10\%$-accuracy domains of $K_{\rm corr}$ and $K_{\rm LD}$ are
separated by a finite region in which neither endpoint is
quantitatively adequate. 
Representative storage-ring parameters span
different regions of the map.
Notably, ESRF-EBS and HEPS, both operating
at 6~GeV, have substantially different
phase-memory parameters, illustrating that beam energy alone
does not determine the applicability of the
limiting theories.
The corresponding machine parameters and references are given in Table I of the End Matter.
The detuning dependence of the validity domains
is examined in the Supplemental Material.

These two requirements imply a nontrivial asymptotic ordering.
A controlled local-density limit is
$|M|\to\infty$, $a\to0$, and $M^2a\to\infty$: the short-lag scale
$|M|^{-1}$ vanishes while
$\Gamma=4\pi M^2a\to\infty$ suppresses synchrotron recurrence.
By contrast, along $a=c/M^2$ with fixed $c$, the short-lag response
becomes asymptotically local but $\Gamma=4\pi c$ remains finite, so
recurrence survives.

In summary, we have established a unified
finite-memory description of radiative depolarization
near an isolated spin resonance.
The connected phase-memory response determines
the leading off-center weak-resonance depolarization
rate and recovers the conventional correlated-sideband
and local resonance-density results as controlled
asymptotic limits.
The crossover is governed by surviving synchrotron
phase memory: localization of the short-lag response
alone does not establish the local-density limit
unless subsequent recurrences are also suppressed.
Monte Carlo spin tracking validates the finite-memory
prediction, including its generalized-sideband structure,
while the resulting validity map provides quantitative
criteria for the applicability of the limiting theories.
This formulation offers a basis for extending radiative
depolarization theory to more general stochastic
spin-orbit dynamics in storage rings.

\begin{acknowledgments}
This work was supported by the National Natural Science Foundation of China (Grant No.~12675189).
\end{acknowledgments}

\bibliography{prl_phase_memory}

@article{SokolovTernov1964,
  author  = {A. A. Sokolov and I. M. Ternov},
  title   = {On Polarization and Spin Effects in the Theory of
             Synchrotron Radiation},
  journal = {Sov. Phys. Dokl.},
  volume  = {8},
  pages   = {1203},
  year    = {1964}
}

@article{DerbenevKondratenko1973,
  author  = {Ya. S. Derbenev and A. M. Kondratenko},
  title   = {Polarization Kinetics of Particles in Storage Rings},
  journal = {Sov. Phys. JETP},
  volume  = {37},
  pages   = {968--973},
  year    = {1973}
}

@article{DKS1979,
  author  = {Ya. S. Derbenev and A. M. Kondratenko and A. N. Skrinsky},
  title   = {Radiative Polarization at Ultrahigh Energies},
  journal = {Particle Accelerators},
  volume  = {9},
  pages   = {247--266},
  year    = {1979}
}

@article{Mane1987,
  author  = {S. R. Mane},
  title   = {Electron-Spin Polarization in High-Energy Storage Rings.
             {I}. Derivation of the Equilibrium Polarization},
  journal = {Phys. Rev. A},
  volume  = {36},
  pages   = {105--119},
  year    = {1987},
  doi     = {10.1103/PhysRevA.36.105}
}

@inproceedings{Assmann1999,
  author    = {R. W. Assmann},
  title     = {The Regimes of Polarization in a High Energy
               $e^+e^-$ Storage Ring},
  booktitle = {Proceedings of the 1999 Particle Accelerator Conference},
  pages     = {3002--3004},
  year      = {1999}
}

@article{BarberEllisonHeinemann2004,
  author  = {D. P. Barber and J. A. Ellison and K. Heinemann},
  title   = {Quasiperiodic spin-orbit motion and spin tunes in storage rings},
  journal = {Phys. Rev. ST Accel. Beams},
  volume  = {7},
  pages   = {124002},
  year    = {2004},
  doi     = {10.1103/PhysRevSTAB.7.124002}
}

@inproceedings{Assmann2001,
  author    = {R. Assmann and J. Badier and A. Blondel and
               M. B{\"o}ge and M. Crozon and B. Dehning and
               H. Grote and J. P. Koutchouk and M. Placidi and
               R. Schmidt and F. Sonnemann and F. Tecker and
               J. Wenninger},
  title     = {Spin Dynamics in {LEP} with 40--100 {GeV} Beams},
  booktitle = {AIP Conference Proceedings},
  volume    = {570},
  pages     = {169--178},
  year      = {2001},
  doi       = {10.1063/1.1384062}
}

@article{Xia2023,
  author  = {Wenhao Xia and Zhe Duan and Desmond P. Barber and
             Yiwei Wang and Bin Wang and Jie Gao},
  title   = {Evaluation of radiative depolarization in the future
             circular electron-positron collider},
  journal = {Phys. Rev. Accel. Beams},
  volume  = {26},
  number  = {9},
  pages   = {091001},
  year    = {2023},
  doi     = {10.1103/PhysRevAccelBeams.26.091001}
}

@article{CEPCTDR2024,
  author  = {{CEPC Study Group}},
  title   = {{CEPC} {Technical Design Report: Accelerator}},
  journal = {Radiat. Detect. Technol. Methods},
  volume  = {8},
  pages   = {1--1105},
  year    = {2024},
  doi     = {10.1007/s41605-024-00463-y},
  eprint  = {2312.14363},
  archivePrefix = {arXiv},
  primaryClass  = {physics.acc-ph}
}

@article{Arnaudon1995,
  author  = {L. Arnaudon and R. Assmann and A. Blondel and
             B. Dehning and P. Grosse-Wiesmann and R. Jacobsen and
             M. Jonker and J. P. Koutchouk and J. Miles and
             R. Olsen and M. Placidi and R. Schmidt and J. Wenninger},
  title   = {Accurate determination of the {LEP} beam energy by
             resonant depolarization},
  journal = {Z. Phys. C},
  volume  = {66},
  pages   = {45--62},
  year    = {1995},
  doi     = {10.1007/BF01496579}
}

@article{BlondelGianfelice2021,
  author  = {Alain Blondel and Eliana Gianfelice},
  title   = {The challenges of beam polarization and keV-scale
             centre-of-mass energy calibration at the {FCC-ee}},
  journal = {Eur. Phys. J. Plus},
  volume  = {136},
  pages   = {1103},
  year    = {2021},
  doi     = {10.1140/epjp/s13360-021-02038-y}
}

@article{Thomas1926,
  author  = {L. H. Thomas},
  title   = {The Motion of the Spinning Electron},
  journal = {Nature},
  volume  = {117},
  pages   = {514},
  year    = {1926},
  doi     = {10.1038/117514a0}
}

@article{BargmannMichelTelegdi1959,
  author  = {V. Bargmann and L. Michel and V. L. Telegdi},
  title   = {Precession of the Polarization of Particles Moving in a
             Homogeneous Electromagnetic Field},
  journal = {Phys. Rev. Lett.},
  volume  = {2},
  pages   = {435--436},
  year    = {1959},
  doi     = {10.1103/PhysRevLett.2.435}
}

@article{Kondratenko1974,
  author  = {A. M. Kondratenko},
  title   = {The Stability of the Polarization of Colliding Beams},
  journal = {Sov. Phys. JETP},
  volume  = {39},
  pages   = {592--595},
  year    = {1974}
}

@article{Yokoya1983Chromaticity,
  author  = {K. Yokoya},
  title   = {Spin Chromaticity for Higher-Order Synchrotron Resonances},
  journal = {Particle Accelerators},
  volume  = {13},
  pages   = {85--93},
  year    = {1983}
}

@article{Mane2005,
  author  = {Mane, S. R. and Shatunov, Yu. M. and Yokoya, K.},
  title   = {Spin-polarized charged particle beams in high-energy accelerators},
  journal = {Rep. Prog. Phys.},
  volume  = {68},
  pages   = {1997--2265},
  year    = {2005},
  doi     = {10.1088/0034-4885/68/9/R01}
}

@inproceedings{Koop2018,
  author    = {Koop, I. A.},
  title     = {Resonant Depolarization at {Z} and {W} at {FCC-ee}},
  booktitle = {Proceedings of the 62nd ICFA Advanced Beam Dynamics Workshop on High Luminosity Circular $e^+e^-$ Colliders (eeFACT2018)},
  address   = {Hong Kong, China},
  pages     = {165--168},
  year      = {2018},
  publisher = {JACoW Publishing},
  doi       = {10.18429/JACoW-eeFACT2018-TUPBB04}
}

@article{FroissartStora1960,
  author  = {M. Froissart and R. Stora},
  title   = {D{\'e}polarisation d'un faisceau de protons polaris{\'e}s dans un synchrotron},
  journal = {Nucl. Instrum. Methods},
  volume  = {7},
  number  = {3},
  pages   = {297--305},
  year    = {1960},
  doi     = {10.1016/0029-554X(60)90033-1}
}

@article{DK1975,
  author  = {Ya. S. Derbenev and A. M. Kondratenko},
  title   = {Relaxation and the equilibrium state of polarization of electrons in storage rings},
  journal = {Sov. Phys. Dokl.},
  volume  = {19},
  pages   = {438--439},
  year    = {1975}
}

@article{BaierOrlov1966,
  author  = {V. N. Baier and Yu. F. Orlov},
  title   = {Quantum depolarization of electrons in a magnetic field},
  journal = {Sov. Phys. Dokl.},
  volume  = {10},
  pages   = {1145},
  year    = {1966}
}

@phdthesis{Sonnemann1998,
  author = {F. Sonnemann},
  title  = {Increase of Spin Polarization for Energy Calibration at {LEP}},
  school = {RWTH Aachen University},
  year   = {1998},
  note   = {{CERN-THESIS-98-014}, PITHA-98-8}
}

@article{LEPEnergy2005,
  author  = {R. Assmann and E. Barbero Soto and D. Cornuet and
             B. Dehning and M. Hildreth and J. Matheson and
             G. Mugnai and A. M{\"u}ller and E. Peschardt and
             M. Placidi and J. Prochnow and F. Roncarolo and
             P. Renton and E. Torrence and P. S. Wells and
             J. Wenninger and G. Wilkinson},
  title   = {Calibration of Centre-of-Mass Energies at {LEP 2} for a
             Precise Measurement of the {W} Boson Mass},
  journal = {Eur. Phys. J. C},
  volume  = {39},
  pages   = {253--292},
  year    = {2005},
  doi     = {10.1140/epjc/s2004-02108-8}
}

@article{BlondelZimmermann2012,
  author        = {Alain Blondel and Frank Zimmermann},
  title         = {A High Luminosity $e^+e^-$ Collider in the {LHC}
                   Tunnel to Study the {Higgs} Boson},
  journal       = {arXiv:1112.2518,},
  year          = {2012},
  doi           = {10.48550/arXiv.1112.2518}
}

@techreport{ESRFEBSDesign2018,
  author      = {{European Synchrotron Radiation Facility}},
  title       = {{EBS Storage Ring Technical Report}},
  institution = {European Synchrotron Radiation Facility},
  address     = {Grenoble, France},
  year        = {2018},
  url         = {https://www.esrf.eu/files/live/sites/www/files/about/upgrade/documentation/Design%20Report-reduced-jan19.pdf},
  note        = {Table 1}
}

@article{XuHEPS2023,
  author  = {H. Xu and C. Meng and Y. Peng and S. Tian and N. Wang
             and X. Cui and C. Du and Z. Duan and others},
  title   = {Equilibrium Electron Beam Parameters of the {High Energy
             Photon Source}},
  journal = {Radiat. Detect. Technol. Methods},
  volume  = {7},
  pages   = {279--287},
  year    = {2023},
  doi     = {10.1007/s41605-022-00374-w}
}

@book{Gardiner2009,
  author    = {Crispin W. Gardiner},
  title     = {Stochastic Methods: A Handbook for the Natural
               and Social Sciences},
  edition   = {4},
  publisher = {Springer},
  address   = {Berlin},
  year      = {2009}
}

@article{Yokoya1983SpinFlip,
  author  = {K. Yokoya},
  title   = {Effects of Radiative Diffusion on the Spin-Flip
             in Electron Storage Rings},
  journal = {Particle Accelerators},
  volume  = {14},
  pages   = {39--61},
  year    = {1983}
}

@article{Blinov2022,
  author  = {V. E. Blinov and E. B. Levichev
             and S. A. Nikitin and I. B. Nikolaev},
  title   = {Resonant Depolarization Technique at
             {VEPP-4M} in Novosibirsk},
  journal = {European Physical Journal Plus},
  volume  = {137},
  pages   = {717},
  year    = {2022},
  doi     = {10.1140/epjp/s13360-022-02825-1}
}

% ================================================================
% END MATTER
% ================================================================

\section*{End Matter}

\paragraph{Second-order phase-memory response.---}
The relation between phase memory and spin relaxation
can be derived directly from the single-particle
spin dynamics.
Using $\dif p=x\,\dif\psi$, the exact rotation
$\widetilde{\bm S}=R_z(-Mp)\bm S$
transforms Eq.~(\ref{eq:spin}) into
\begin{equation}
\dot{\widetilde{\bm S}}
=
[\bm\Omega_c+U\bm\eta(p)]
\times\widetilde{\bm S},
\label{eq:EM_rotated_spin}
\end{equation}
where
$\bm\Omega_c=U\mu\bm e_x+\delta\bm e_z$,
$\mu=e^{-M^2/2}$, and
$\bm\eta(p)=
[\cos(Mp)-\mu]\bm e_x-\sin(Mp)\bm e_y$.
The coherent mean spin-precession field
produces deterministic precession, while
its zero-mean fluctuations generate
spin relaxation through their time correlations.

Choose a right-handed basis
$(\bm e_1,\bm e_2,\bm e_3)$ with
$\bm e_3=\bm\Omega_c/\Omega_c$,
$\Omega_c=|\bm\Omega_c|$.
Writing
$S_+=\widetilde S_1+i\widetilde S_2$,
$S_3=\widetilde{\bm S}\cdot\bm e_3$,
and $\eta_+=\eta_1+i\eta_2$,
the single-particle equations become
\begin{equation}
\begin{aligned}
\dot S_+
&=i(\Omega_c+U\eta_3)S_+
-iU\eta_+S_3,\\
\dot S_3
&=U\,\operatorname{Im}(\eta_+^*S_+).
\end{aligned}
\label{eq:EM_spin_components}
\end{equation}
Here $\eta_j=\bm\eta\cdot\bm e_j$.
The fluctuating transverse field first
generates a spin response of order $U$,
which feeds back into $S_3$ at order $U^2$.

For an initially coherent-axis-aligned spin,
$S_+(0)=0$ and $S_3(0)=1$.
Solving the transverse equation by an
integrating factor and substituting into
the longitudinal equation gives the exact
single-particle identity
\begin{equation}
\begin{aligned}
\dot S_3(\psi)
={}&-U^2\operatorname{Re}
\int_0^\psi
E(\psi,v)
\eta_+^*(\psi)\eta_+(v)\\
&\qquad\times S_3(v)\,\dif v,
\end{aligned}
\label{eq:EM_two_field}
\end{equation}
where
$E(\psi,v)=
\exp\{i\Omega_c(\psi-v)
+iU\int_v^\psi\eta_3(w)\,\dif w\}$.
This identity shows that the longitudinal
spin evolution depends on the two-time
response of the fluctuating spin-precession field.

For a stationary orbital ensemble, define
$P_\parallel(\psi)=\langle S_3(\psi)\rangle$.
At fixed finite time, the weak-$U$ expansion
gives $S_3(v)=1+O(U^2)$ and
$E(\psi,v)=e^{i\Omega_c(\psi-v)}+O(U)$.
Substituting these into
Eq.~(\ref{eq:EM_two_field}) and averaging
over the orbital ensemble gives
$\dot P_\parallel(\psi)
=-U^2\int_0^\psi
C_{\rm spin}(\vartheta)\,\dif\vartheta
+o(U^2)$,
where stationarity has been used to express
the response in terms of the lag
$\vartheta=\psi-v$.
The projected second-order correlation is
\begin{equation}
\begin{aligned}
C_{\rm spin}(\vartheta)
&=
\operatorname{Re}
\left[
e^{i\Omega_c\vartheta}
\left\langle
\eta_+^*(\psi+\vartheta)
\eta_+(\psi)
\right\rangle
\right]\\
&\longrightarrow
\operatorname{Re}G_c(\vartheta)
\quad (U\to0,\ \delta\ne0).
\end{aligned}
\label{eq:EM_projected_kernel}
\end{equation}
At fixed nonzero detuning,
$\Omega_c\to|\delta|$ and the coherent-axis
tilt vanishes as $U\to0$.
Gaussian averaging gives
$\langle\eta_+^*(\psi+\vartheta)
\eta_+(\psi)\rangle
\to e^{-M^2[1-C_p(\vartheta)]}-e^{-M^2}$.
Thus $C_{\rm spin}$ reduces to
$\operatorname{Re}G_c$, establishing the
connection between the second-order spin
response and the connected phase-memory
correlation of Eq.~(\ref{eq:connected_kernel}).

The fixed-time second-order expansion develops
a secular contribution at long times and must
be resummed on the kinetic timescale
$\psi=O(U^{-2})$.
In the weak-$U$ limit at fixed $a$, $M$,
and nonzero $\delta$, the connected memory
is integrable, and the isolated slow spin
mode admits the leading secular response
$\dot P_\parallel(\psi)
=-U^2\int_0^\psi
C_{\rm spin}(\vartheta)
P_\parallel(\psi-\vartheta)
\,\dif\vartheta
+\text{higher-order terms}$.

Substituting the post-transient form
$P_\parallel(\psi)=A e^{-\rho_d\psi}$
gives
\begin{equation}
\begin{aligned}
\rho_d
&=U^2\int_0^\infty
C_{\rm spin}(\vartheta)
e^{\rho_d\vartheta}\,\dif\vartheta
+o(U^2)\\
&=U^2\int_0^\infty
C_{\rm spin}(\vartheta)\,\dif\vartheta
+o(U^2)\\
&=U^2K_\Phi+o(U^2).
\end{aligned}
\label{eq:EM_slow_rate}
\end{equation}
Since $\rho_d=O(U^2)$ and the connected
correlation has finite lag moments,
the exponential factor contributes only
at $O(U^4)$.
The last equality follows from
$C_{\rm spin}(\vartheta)\to
\operatorname{Re}G_c(\vartheta)$
in the fixed-detuning weak-$U$ limit.
This recovers Eq.~(\ref{eq:main}).

The result gives the leading decay exponent
of an isolated slow spin mode, rather than
an exact description of arbitrary polarization
histories.
Initial spin-orbit correlations may modify
transient amplitudes, while coherent resonance
cores require the full vector spin response.

\paragraph{Correlated-sideband limit.---}
In the weak-radiation regime, synchrotron motion
remains nearly periodic, and resonance-driving
spin responses retain phase coherence across successive
synchrotron periods.
Their recurrent interference produces resolved
synchrotron sidebands.
The correlated-sideband result follows by first
taking $U\to0$ at fixed $a>0$, then $a\to0^+$
at fixed $M$ and detuning separated from
integer sidebands.

To control the long-lived recurrences, we use
an exact finite-$a$ representation.
With $z=M^2$, $\omega=\sqrt{1-a^2}$, and
$b=(z/2)(1-ia/\omega)$, the orbital factor satisfies
$zC_p(\vartheta)
=b e^{(-a+i\omega)\vartheta}
+b^*e^{(-a-i\omega)\vartheta}$
for $\vartheta\geq0$.
Expanding $e^{zC_p}$ and integrating term by term
gives
\begin{equation}
K_\Phi
=
\operatorname{Re}\,
e^{-z}
\sum_{\substack{j,k\geq0\\j+k>0}}
\frac{b^j(b^*)^k}{j!\,k!}
\frac{1}{
a(j+k)-i[\delta+(j-k)\omega]
}.
\label{eq:EM_exact_series}
\end{equation}
The omitted $(j,k)=(0,0)$ term is the coherent
mean-product contribution.
Here $j-k$ determines the harmonic frequency,
while $a(j+k)$ gives the width of each pole
contribution.

For detunings bounded away from integer sidebands,
the factorial weights permit a controlled
weak-$a$ expansion.
The real part of the analytically continued
pole series is odd in $a$, eliminating even
powers.
Grouping $j-k=m$ and using the Bessel recurrence
relations gives
$K_\Phi=K_{\rm corr}+O(a^3)$,
recovering Eq.~(\ref{eq:endpoints}).
This expansion is nonuniform near integer
sidebands, where coherent resonance dynamics
must be retained.

The leading coefficient independently agrees
with the conventional radiation-free
ISF-gradient recoil calculation:
$\rho_{d,\rm corr}
=2a\langle|\partial_x\bm n|^2\rangle
=U^2K_{\rm corr}+O(aU^4)$.
Here differentiation is with respect to the
energy coordinate $x$ at fixed $p$, consistent
with photon recoil acting directly on the
energy coordinate.
The perturbative ISF solution and the
normalization calculation are given in the
Supplemental Material.

\paragraph{Local-density limit.---}
The local-density limit follows from the
short-lag reduction of the connected response
in Eq.~(\ref{eq:connected_kernel}).
For $|M|\gg1$, the central contribution is
localized on $\vartheta\sim |M|^{-1}\ll1$.
Expanding the stationary autocorrelation gives
$C_p(\vartheta)
=1-\vartheta^2/2+O(a\vartheta^3,\vartheta^4)$.
Consequently, within the central short-lag region,
\begin{equation}
G_c(\vartheta)
\simeq
e^{i\delta\vartheta-M^2\vartheta^2/2}
=
\left\langle e^{i\Delta\vartheta}\right\rangle,
\label{eq:EM_local_kernel}
\end{equation}
where $\Delta=\delta+Mx$ is the normalized
instantaneous resonance detuning.
The local kernel is therefore the characteristic
function of the frozen detuning, consistent
with $\Delta\Phi\simeq\Delta\vartheta$.
The coherent mean-product subtraction is
exponentially small for $|M|\gg1$.

Substituting this local kernel into the causal
response integral and evaluating the Gaussian
cosine transform gives
\begin{equation}
\begin{aligned}
K_{\rm LD}
&=
\int_0^\infty
e^{-M^2\vartheta^2/2}
\cos(\delta\vartheta)\,\dif\vartheta
\\
&=
\pi p_\Delta(0),
\end{aligned}
\label{eq:EM_KLD}
\end{equation}
recovering the local resonance-density
coefficient in Eq.~(\ref{eq:endpoints}).

This central contribution is the asymptotic
form of the full $K_\Phi$ only when the
integrated contributions from subsequent
synchrotron recurrences are negligible.
The ordering
$|M|\to\infty$, $a\to0$, and $M^2a\to\infty$
ensures $K_\Phi/K_{\rm LD}\to1$
for bounded $\delta/M$.
Thus localization of the central response
and loss of inter-period phase memory
are distinct requirements.

The coefficient $\pi p_\Delta(0)$ also agrees
with the conventional uncorrelated
resonance-passage result~\cite{Kondratenko1974}.
The corresponding weak Froissart--Stora
normalization and the control of the full
recurrence contribution are detailed in
the Supplemental Material.

\begin{table*}[t]
\caption{
Representative longitudinal parameters of storage rings
and their mapping onto Fig.~\ref{fig:regime_map}.
}
\label{tab:machine_mapping}
\begin{ruledtabular}
\begin{tabular}{lccccccc}
Machine
& $E$ (GeV)
& $\sigma_\epsilon$
& $\nu_s$
& $\lambda_z$
& $M$
& $a$
& $\Gamma$ \\
\hline
LEP~\cite{Sonnemann1998,LEPEnergy2005}
& 60.6
& $9.60\times10^{-4}$
& 0.111
& $6.51\times10^{-3}$
& 1.189
& $9.33\times10^{-3}$
& 0.166 \\

LEP~\cite{BlondelZimmermann2012}
& 104.5
& $2.20\times10^{-3}$
& 0.142
& $1.79\times10^{-2}$
& 3.667
& $2.01\times10^{-2}$
& 3.390 \\

CEPC~\cite{CEPCTDR2024}
& 45.5
& $4.00\times10^{-4}$
& 0.035
& $8.18\times10^{-4}$
& 1.180
& $3.72\times10^{-3}$
& 0.0651 \\

CEPC~\cite{CEPCTDR2024}
& 80
& $7.00\times10^{-4}$
& 0.062
& $4.44\times10^{-3}$
& 2.050
& $1.14\times10^{-2}$
& 0.603 \\

CEPC~\cite{CEPCTDR2024}
& 120
& $1.00\times10^{-3}$
& 0.049
& $1.50\times10^{-2}$
& 5.558
& $4.86\times10^{-2}$
& 18.86 \\

ESRF-EBS~\cite{ESRFEBSDesign2018}
& 6
& $9.30\times10^{-4}$
& $3.66\times10^{-3}$
& $3.08\times10^{-4}$
& 3.460
& $1.34\times10^{-2}$
& 2.017 \\

HEPS~\cite{XuHEPS2023}
& 6
& $1.02\times10^{-3}$
& $8.046\times10^{-4}$
& $2.42\times10^{-4}$
& 17.28
& $4.78\times10^{-2}$
& 179.51
\end{tabular}
\end{ruledtabular}

\begin{flushleft}
\footnotesize
The LEP 104.5-GeV point uses representative LEP2
parameters. CEPC energy spreads are natural values,
excluding beamstrahlung.
The HEPS parameters correspond to the bare lattice
with the harmonic cavity off.
\end{flushleft}
\end{table*}

\paragraph{Monte Carlo spin tracking.---}
For Fig.~\ref{fig:phase_memory_validation}, we numerically
integrate the coupled dynamics of
Eqs.~(\ref{eq:spin}) and~(\ref{eq:orbit}).
The longitudinal coordinates are initialized in the stationary
Gaussian ensemble and advanced using the exact finite-step
Gaussian transition~\cite{Gardiner2009}.
The accumulated detuning phase over each step is
$\delta h+M(p_{j+1}-p_j)$, evaluated from the same
orbital realization.
The spin is advanced by a norm-preserving $SO(3)$
Rodrigues rotation with rotation vector
$(Uh,0,\delta h+M[p_{j+1}-p_j])$.
This finite-step map approximates the time-ordered
spin evolution.

Each particle's initial spin is aligned with the
deterministic periodic ISF of the radiation-free
single-resonance system, evaluated at its initial
orbital coordinates:
$\bm S_i(0)=\bm n[x_i(0),p_i(0)]$.  The tracked polarization is
\begin{equation}
 P_n(\psi)=
 \left\langle
 \bm S(\psi)\cdot
 \bm n[x(\psi),p(\psi)]
 \right\rangle .
 \label{eq:EM_tracking_observable}
\end{equation}
Although $P_n$ differs from the coherent-axis
polarization used in the analytical derivation,
both probe the same slow relaxation mode in the
off-center weak-resonance regime.
Their fitted decay rates agree within statistical
uncertainty for representative correlated,
finite-memory, and local-density conditions.

The depolarization rate is extracted from the late-time
exponential decay $P_n(\psi)=A\exp(-\rho_d\psi)$,
with $A$ fitted independently.
All 59 tracking points use $U=0.01$, a spin-integration
step $h=0.05$, eight independent random seeds, and
2048 particles per seed.
The primary fit window is
$\psi\in[5/a,0.90\psi_{\rm end}]$.
Statistical errors are estimated from 320 hierarchical
seed/block bootstrap refits.
Ensemble-size, fit-window, and $U^2$-scaling tests
verify the stability of the extracted rate.

For the local-density benchmark, common-random-number
step refinement on nested exact orbital trajectories
gives near-quadratic convergence.
At the production step size $h=0.05$, the
extrapolated finite-step correction is only
$0.032\%$ of the depolarization rate.
Correlation-aware fit-window tests, including
a delayed fit start at $a\psi=30$, confirm
the stability of the extracted decay rate
against the exclusion of early-time transients.
Points within narrow coherent resonance cores
are excluded from the scalar-rate comparison.

\paragraph{Mapping representative storage rings onto
the phase-memory plane.---}
Representative longitudinal parameters of LEP,
CEPC, ESRF-EBS, and HEPS are mapped onto the
$(M,\Gamma)$ plane in Fig.~\ref{fig:regime_map}.
The parameters and their sources are listed in
Table~\ref{tab:machine_mapping}.
The mapping assumes planar integer spin resonances,
for which $M\simeq\nu_0\sigma_\epsilon/\nu_s$.
These points characterize longitudinal phase-memory
conditions rather than complete polarization lifetimes,
which also depend on resonance strengths, detunings,
and the full spin-orbit dynamics.

\end{document}